\documentclass[10pt]{iopart}
\usepackage{epsfig}
\begin{document}

\title[Critical Pressure]{Critical Hadronization Pressure}

\author{Johann Rafelski$^\dag$  and 
Jean Letessier$^\ddag$}
\address{
\dag\ Department of Physics, 
University of Arizona, Tucson, AZ 85721, USA\\
\ddag\  LPTHE,
Universit\'e Paris 7, 2 place Jussieu, F--75251 Cedex 05}

\begin{abstract}
We discuss the bulk properties of QGP produced at RHIC obtained at
time of hadronization. We argue that
 hadronization of quark--gluon plasma occurs at 
a  critical pressure near to $82$ MeV/fm$^3$, obtained for SPS energy range. 
\end{abstract}

%\pacs{12.38.Mh}

%\submitto{\JPG}

\section{Introduction}
We study  the physical properties of the hot and dense 
fireball, and in particular,   the thermal pressure 
at time of hadronization. In our approach,  we
consider  that   at RHIC, as well as 
at the top SPS energies,  a local domain of thermally colliding   gluons and quarks 
has been formed, the  quark--gluon plasma (QGP).  In the early stages
of the collision   a high density, and thus pressure buildup occurs 
 which is followed by  a fast  relativistic 
collective matter outflow. This system  expansion dilutes the density 
beyond  phase equilibrium transformation to hadron gas (HG). 

This sub-dense  system is unstable and
can experience a sudden  breakup, converting rapidly from quarks 
into hadrons, with  free streaming particles, only `strong' hadron 
resonances are subject to possible further interactions. However, this does
not alter the final stable particle yields. Thus, fitting the yields of 
particles using  
the statistical hadronization model (SHM),  we can infer from the   measured 
hadronic particle yields also the yields of other unmeasured hadrons. Summing
the contributions of many gas fractions, we
obtain the physical conditions of the fireball. 

In such a procedure, it is necessary to allow for greatest possible flexibility
in characterization of particle phase space, consistent with conservation 
laws and related physical constraints  of QGP breakup. In particular,  
the QGP yield of strange and light quark pairs has to be   nearly preserved while
QGP particles are  distributed into final state hadrons. This is accomplished  
using parameters which describe the quark pair yields, i.e., $\gamma_q$ and $ \gamma_s$.
While these parameters in both phases are not equal, 
i.e., $\gamma_i^{\rm QGP}\ne \gamma_i^{\rm HG},\ i=q,\ s$, 
the yield of pairs are similar, i.e., $N_i^{\rm QGP}\simeq N_i^{\rm HG}$, provided that
 the QGP breakup process is rapid, and that is seen in the HBT data.  A value of 
$ \gamma_i \ne 1$ allows to control the density of particles at given temperature,
but unlike the chemical potential, $ \gamma_i$ acts in the same direction for particles 
and conjugate (anti) particles. How this works will become clear in next section.

A  jump-up in the phase
 space occupancy parameter  $\gamma_q$  replaces an increase in volume associated with  a 
slow re-equilibrating hadronization  with  mixed phase, a reaction picture incompatible with 
many reaction observables, including HBT.
The rise in occupancy, just like the rise in the volume size needed
when chemical  equilibrium $\gamma_i^{\rm HG}=1$ is assumed,
accommodates transformation of a  entropy
dense QGP  phase into  entropy dilute HG  phase.
Similarly, a  jump-up in the strangeness phase
 space occupancy parameter  $\gamma_s$  allows for higher density of  strangeness in
QGP compared to HG. We have, in general,
$$\gamma_i^{\rm HG}(t_f)  > \gamma_i^{\rm QGP}(t_f),\ i=q,\ s.$$

The available number of quark pairs in QGP at hadronization decisively influences the 
possibility to form baryons. In fact, the  {\em ratio} of baryon to meson yield arising in 
 microscopic dynamics of hadronization is proportional to $\gamma_q$.  This 
 establishes  the necessity to include the occupancy parameters in 
 order to describe the yields of hadrons, since this is the parameter which 
allows   for a hadronization dependent dynamical 
relative yield of mesons and baryons. 
Conversely,  a study of particle yields with a 
fixed  light quark equilibrium value $\gamma_q^{\rm HG}=1$   
presumes that the relative yield of baryons to mesons 
is fully chemically equilibrated, and that we know well 
the spectrum of hadrons. Clearly, neither assumption is safe and 
the choice $\gamma_q^{\rm HG}=1$ 
is  over-constraining any hadronization model, as of course is the
choice $\gamma_s^{\rm HG}=1$.

%%%%%%%%%%%%%%%%%%%%%%%%%%%%%%%%%%%%%%%%%%%%%%%%%%%%%%%%%%%%%%%
\section{Particle yield SHM data analysis}
 
The analysis of experimental hadron yield results 
requires a significant book-keeping and fitting  effort in 
order to allow for resonances, particle
widths, full decay trees and isospin multiplet sub-states.
A program SHARE (Statistical HAdronization with REsonances)
suitable to perform this data analysis 
is available for public use~\cite{Torrieri:2004zz,Torrieri:2006xi}.
This program implements the PDG~\cite{Yao:2006px} confirmed (4-star) 
set of particles and resonances, and we use~\cite{Letessier:2005qe}
already for two years the modern $\sigma$-meson mass~\cite{Yndurain:2007qm}
($m_\sigma=484,\ \Gamma_\sigma/2=255$ MeV).

The important parameters
of the SHM, which control the relative yields
of particles, are the particle specific fugacity factors ${\lambda}$ and the
space occupancy factors ${\gamma}$ discussed above. The fugacity is related to chemical
potential ${\mu} = T{\ln{\lambda}}$. The occupancy   ${\gamma}$ is, nearly,
the ratio of produced   particles to the number of particle
expected in chemical equilibrium, and thus, meson yield is (nearly) proportional to 
$\gamma^2$ and baryon yield to $\gamma^3$ (here, we did not distinguish the valance
quark content for $u,\,d,\,s$ quarks). The actual formula for the momentum  distribution is,
both for the HG and QGP phase:
\begin{equation}
{{d^6N}\over {d^3pd^3x}}\equiv f(p)={g\over (2\pi)^3}
               {1\over \gamma^{-1}\lambda^{-1} e^{E/T}\pm 1}
             \to \gamma\lambda e^{-E/T},
\end{equation}
where the Boltzmann limit of the Fermi `$(+)$' and Bose `$(-)$' distributions, applicable, 
in particular when $m/T>1$, is indicated.
$g$ is the degeneracy factor, $T$ is the temperature and $E=E(p)$ is the single particle 
energy spectrum, typically $E=\sqrt{m^2+p^2}$.

 The fugacity   ${\lambda}$ is associated with a
conserved quantum number, such as  net-baryon number, net-strangeness or heavy flavor.
Thus, antiparticles
have inverse value of  ${\lambda}$, and  ${\lambda}$  evolution during
the reaction process is related to the
changes in   densities due to dynamics, such as expansion. Contrary to  ${\lambda}$,
 ${\gamma}$  is the same for particles and antiparticles.  Its value
changes as a function of time, even if the system does not expand, 
for it describes buildup in time of
the particular particle species.
For this reason, ${\gamma}$ changes rapidly during the reaction,
while ${\lambda}$ is more constant. It is ${\gamma}$ which carries
the information about the time history of the reaction and the precise
condition of particle production referred to as chemical freeze-out.

In the quark phase, each particle has its proper chemical  yield co-factor, thus 
for light quarks $q=u,\,d$, we have yield co-factors  $\gamma_q^{\rm QGP}\lambda_q^{\rm QGP}$,  
and for antiquarks
 $\bar q$, we have $\gamma_q^{\rm QGP}\lambda_q^{-1\,\rm QGP}$, and similarly
for strange quarks $s$ and antiquarks $\bar s$. In the HG phase, we need to count valance 
quark content of each hadron. For example, for the $\Lambda$, the chemical co-factor 
is $\gamma_s^{\rm HG}\gamma_q^{2\,\rm HG}\lambda_s^{\rm HG}\lambda_q^{2\,\rm HG}$, while for
 $\overline{\Lambda}$, it is  
$\gamma_s^{\rm HG}\gamma_q^{2\,\rm HG}\lambda_s^{-1\,\rm HG}\lambda_q^{-2\,\rm HG}$.
We recall that the chemical potentials of baryon number, $\mu_{\rm B}$, and 
hyperon number, $\mu_{\rm S}$, are
$$
\mu_{\rm B}=3T\ln \lambda_q,\qquad \mu_{\rm S}=T\ln \lambda_q-T\ln \lambda_s,
$$
where, for historical reasons, hyperon number has opposite quantum number to strangeness.
Above and from now on, when the upper index is 
absent the (chemical) variable considered refers to the final state phase, thus 
to  HG.  SHARE allows the conservation of  (electrical) 
charge $Q$, which is done at the cost of introducing the fugacity 
 $\lambda_{I3}\equiv \lambda_u/\lambda_d$ and we note, in this context,  
that  $\lambda_q=\sqrt{\lambda_u\lambda_d}$.

%%%%%%%%%%%%%%%%%%%%%%%%%%%%%%%%%%%%%%%%%%%%%%%%%%%%%%%%%%%% FIG 1
\begin{figure}[tb]  
\centerline{\psfig{width=6cm,figure= 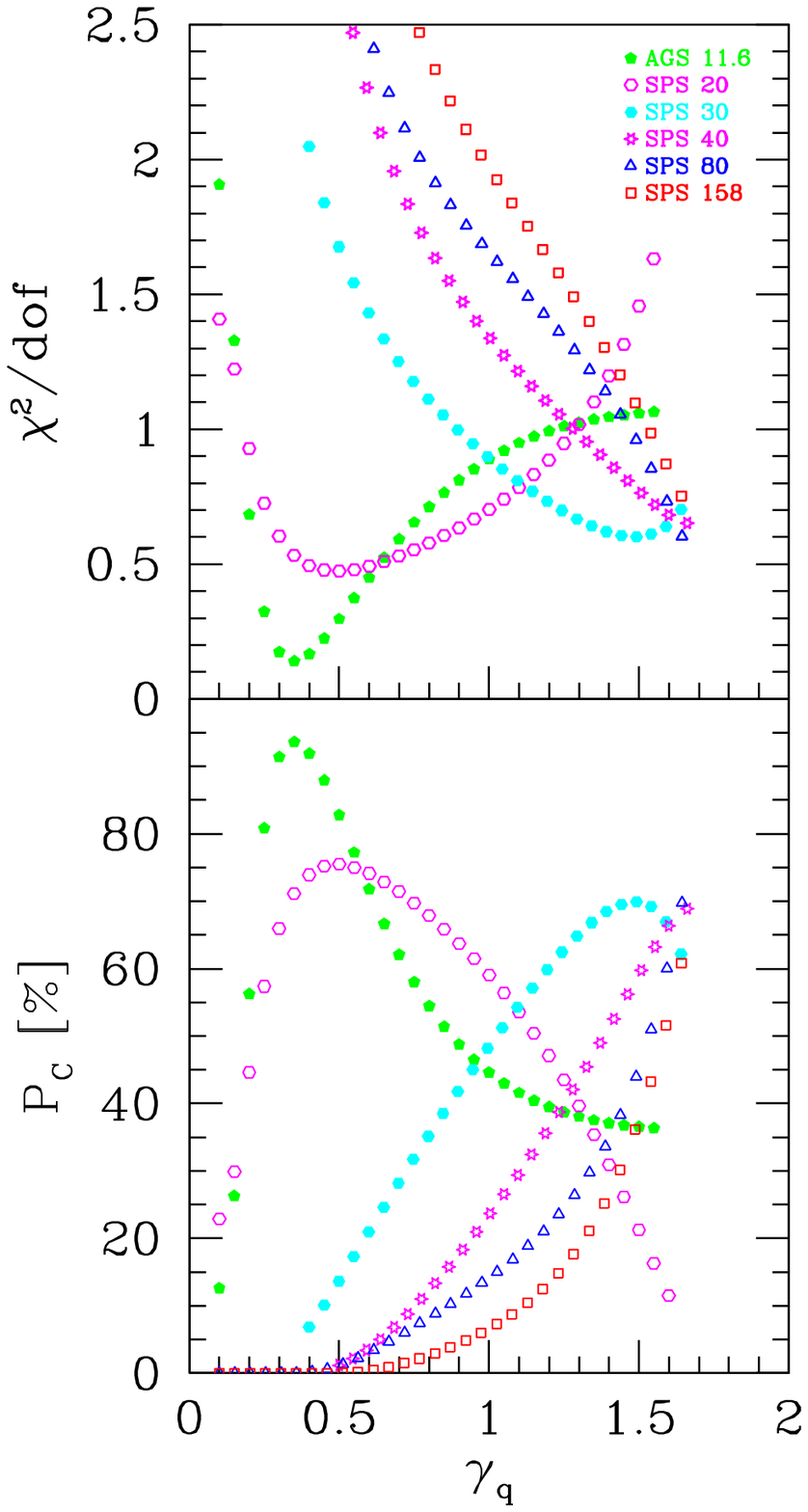}\hspace*{-0.5cm}
            \psfig{width=6cm,figure=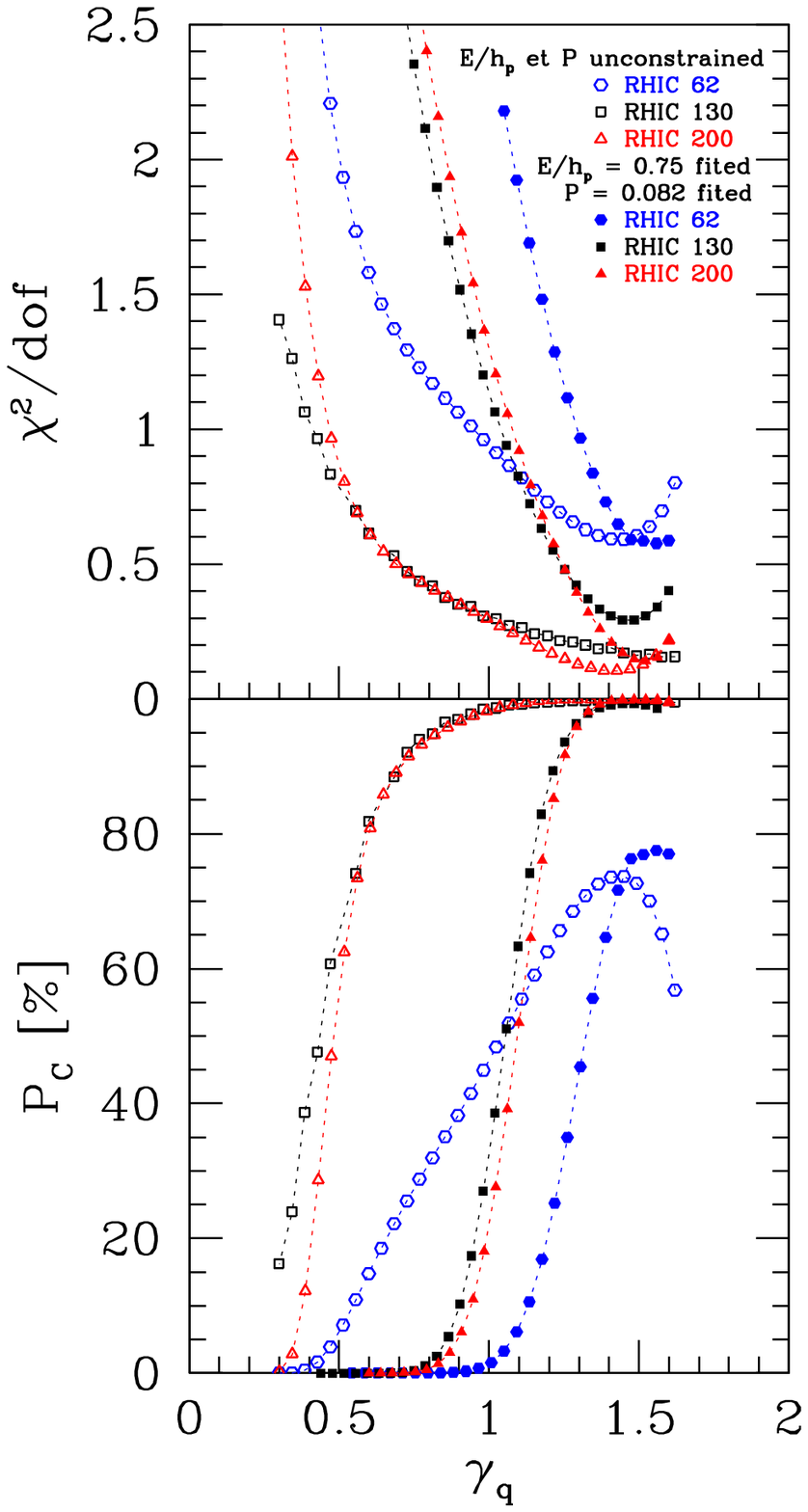}}
\caption{\label{ChiP}
$\chi^2/{\rm dof}$ (top) and the associated  significance level  $P_C[\%]$
(bottom) as function of $\gamma_q$, on left for the AGS/SPS energy range,
and on right for  RHIC.
}
\end{figure} 
%%%%%%%%%%%%%%%%%%%%%%%%%%%%%%%%%%%%%%%%%%%%%%%%%%%%%%%%%%%%%%%%%%%%

We evaluate the success of our data  fit considering 
the profile   of $\chi^2/{\rm dof}$ 
as function of $\gamma_q$. (see top of figure~\ref{ChiP}), 
on left for AGS--SPS energy range, and on right for RHIC. The best
fit is clearly not obtained at $\gamma_q=1$ %(super script `HG' omitted). 
Note that for a small number of
dof, the value of  $\chi^2/{\rm dof}$ can be very misleading, for this reason
the bottom frames in   figure~\ref{ChiP}   show 
the confidence level $P_C\equiv CL$. 

The meaning of  $P_C\equiv CL$ is explained   in 
``Review of Particle Physics''~\cite{Yao:2006px} where we  see, in Figure~32.2, lines of 
 fixed value  of $P_C$, for given values $\chi^2\!/{\rm dof}$ and dof. 
 Note that the value $P_C=50\%$ is equivalent  to
$\chi^2/{\rm dof}=1$ for the case that very many dof are present.
However, for a small number of dof, a very much smaller 
value of  $\chi^2\!/{\rm dof}$ must be achieved to claim good 
confidence fit. Only $P_C(\chi^2\!,\mathrm{dof})$,  
and not  $\chi^2\!/{\rm dof}$, expresses 
confidence in  the validity of the model used to fit the data.

Effectively,  $P_C(\chi^2\!,\mathrm{dof})$ 
 also expresses confidence in the data, provided
that we believe in the model. This is easily recognized by
checking what happens  when we intentionally 
alter an experimental data point,  e.g., by 2 s.d.. We find that our data fit remains
stable in the sense that we find nearly the same model parameters, but $P_C$
becomes much smaller, and the falsified data point  contributes dominantly 
to the error of the fit.  

At SPS (left side, bottom frame of figure~\ref{ChiP}), 
our fits, carried out with the   NA49 2008 data set, converge 
to a common best value  $P_C=70\%$~\cite{Rafelski:2009gu}.
This value has somewhat higher  $P_C$ than one should 
expect on statistical grounds (50\%). However, 
we have treated systematic errors as if these were statistical errors, and have 
added these linearly to the statistical errors, in effect making  many measurement 
errors too large.
On the right hand side, in figure \ref{ChiP}, we see RHIC results. It seems that 
for $\sqrt{s_{\rm NN}}=200$ and 130 GeV, the confidence level is here too high, suggesting
that the combination of statistical errors with systematic errors was not appropriate
(our, here presented, 62 GeV fit comprises some extrapolated and interpolated results 
and should  not be  seen as yet to be a `real' data fit).

%%%%%%%%%%%%%%%%%%%%%%%%%%%%%%%%%%%%%%%%%%%%%%%%%%%%%%%%%%%%%%%%%%%%%%%%
\section{Data used and statistical parameters}
The   data sets for AGS--SPS study were presented elsewhere~\cite{Rafelski:2009gu},
the study of total (as opposed to $dN/dy$) 
hadron yields at RHIC is  relying on extrapolations, and the discussion of this 
matter goes  beyond the scope
of this presentation. In table~\ref{Rhic}, we show the in--out data fields
for the fits  to central rapidity 200, 130 and 62.4 GeV at RHIC, as used here.
As seen in table \ref{Rhic}, we 
combine   PHENIX data for direct single particle spectra with
RHIC date for (strange) particles  yields reconstructed from 
the decay particles 
using, e.g., the invariant mass method~\cite{Timmins}, a more complete
discussion of our data set goes beyond the scope of this report, 
see also~\cite{Letessier:2005qe}.

%%%%%%%%%%%%%%%%%%%%%%%%%%%%%%%%%%%%%%%%%%%%%%%%%%%%%%%%%%%%%%%%%%%%%%%%%%%%%%%%%%%%
\begin{table}[t]
\caption{
\label{Rhic}
The constrains, imposed and natural at the top, followed by input particle data  and 
the resulting statistical parameters,  and at bottom,
the chemical potentials derived from these, 
for   RHIC central rapidity, most central collisions. $\lambda_s$  values are
obtained from the  constraint to zero strangeness. 
The weak feeds allowed were as stated by experimental groups
or/and estimated by us, however, in the fit of the $p, \bar p$ 
yields  we accepted complete weak feed  from all hyperons.
}\vspace*{0.2cm}\small   
\begin{center}
\begin{tabular}{|c| c c c |  }
\hline
$\sqrt{s_{\rm NN}}$  [GeV]  & 62.4          & 130              & 200  \\
%$E_{\rm eq}$ [$A$\,GeV]     &2075 & 9008  &  21321 \\
%$\Delta y$ &$\pm 4.2$  & $\pm 4.93$ & $\pm 5.36$ \\
%\hline
% & \multicolumn{3}{|c|}{ $dN/dy|_{y=0}$\ 5\% }\\
 \hline
$P$ [GeV/fm$^3$]            &0.082$\pm$0.001& 0.082$\pm$0.001 & 0.082$\pm$0.001 \\
$E/{h}_p$ [GeV]             &0.75$\pm$0.075 &0.75$\pm$0.075   & 0.75$\pm$0.075     \\
\hline
$(s-\bar s)/(s+\bar s)$     &  0   &   0   &  0    \\
$Q/b $                      &0.39$\pm$0.01  &0.4$\pm$0.01     &0.4$\pm$0.01 \\
\hline
$\pi^+$                     & 233$\pm$26    & 276$\pm$36      & 286.4$\pm$24.2   \\
$\pi^-$                     & 237$\pm$27    &270$\pm$36       &281.8$\pm$22.8  \\
${\rm K}^+$                 & 38$\pm$4.3    &46.7$\pm$8       &48.9$\pm$6.3      \\
${\rm K}^-$                 & 32.6$\pm$4.7  &40.5$\pm$7       &45.7$\pm$5.2  \\
%$\phi $                    &               &                 &                 \\
$\phi/{\rm K}^-$            &               &0.15$\pm$0.03    &0.174$\pm$0.03 \\
$p$                         &34.3$\pm$3.8   &28.7$\pm$4       &28.3$\pm$4.8  \\
$\bar p$                    &13.8$\pm$1.6   &20.1$\pm$2.8     &13.5$\pm$1.8  \\
%$\bar p/p$                 &               &0.70$\pm$0.06    &0.747$\pm$116  \\
%${\rm K}^+/p$              &               &2.4$\pm$0.5      &                \\
$\Lambda$                   &               &17.35$\pm$0.8    &16.7$\pm$1.3     \\
$\overline\Lambda$          &               &12.5$\pm$0.8     &12.7$\pm$1.1    \\
$\Xi^-$                     &1.84$\pm$0.2   &                  &2.17$\pm$0.25                 \\
$\overline\Xi^+$            & 1.16$\pm$0.12 &                  &1.83$\pm$0.25                 \\
$\Xi^-$/$h^-$               &               &0.0077$\pm$0.0016 &                 \\
$\overline\Xi^+$/$\Xi^-$    &               &0.853$\pm$0.1     &                 \\
$\Omega$                    &0.229$\pm$0.035&                  &                 \\
$\overline\Omega$           &0.176$\pm$0.030&                  &               \\
$\Omega+\overline\Omega$     &              &                   & 0.85$\pm$0.08              \\
$(\Omega+\overline\Omega)/h^-$&              &0.0021$\pm$0.0008&               \\
%$\Omega/h^-$                &               &0.0012$\pm$0.0005 &                 \\
%$\overline\Omega$           &               &                 &               \\
%$\overline\Omega/\Omega$    &               &0.95$\pm$0.01    &               \\
%${\rm K}_{\rm S}$/${\rm K}^-$&              &                 &           \\
%${\rm K}(892)0  $           &               &                 &               \\
${\rm K}^0(892)/{\rm K}^-$   &               &0.26$\pm$0.08     & 0.23$\pm$0.05      \\
 \hline
$dV/dy$ ${\rm [fm}^3]$      &1089$\pm$74      &1172$\pm$93        &1156$\pm$88      \\
$T$ [MeV]                   &142.9$\pm$0.3    &140.2$\pm$0.2    & 140.5$\pm$0.5     \\
$\lambda_q$                 &1.166$\pm$0.036  &1.077$\pm$0.020  &1.066$\pm$0.030\\
$\lambda_s$                 &1.066$^*$        &1.029$^*$        & 1.033$^*$     \\
$\gamma_q$                  &1.53$\pm$0.10    &1.56$\pm$0.031   & 1.54$\pm$0.14     \\
$\gamma_s$                  &1.74$\pm$0.22    & 2.32$\pm$0.31   &2.30$\pm$0.40      \\
$\lambda_{I3}$              &0.992$\pm$0.003  &0.997$\pm$0.001  & 0.997$\pm$0.002 \\
 \hline 
$\mu_B$ [MeV]               &65.9             &31.2             &27.1       \\
$\mu_S$ [MeV]               &13.5             &6.4              &5.6      \\
 \hline
\end{tabular}\vspace*{0.1cm}
\end{center}
 \end{table}
%%%%%%%%%%%%%%%%%%%%%%%%%%%%%%%%%%%%%%%%%%%%%%%%%%%%%%%%%%%%%%%%%%%%%%%%%%%%%%%%%%%%%

Other statistical hadronization (SH) parameters we derive from  the data, shown in bottom
of table \ref{Rhic}, are
the source volume $V$ (that is $dV/dy$ for RHIC, the volume associated with the 
interval of rapidity in which particles are measured), the temperature  $T$, at which 
particles stop changing in yield (chemical freeze-out).  Moreover, we obtain
chemical potentials $\mu_B=3\mu_q=3T\ln \lambda_q,\ \mu_S=  T\ln (\lambda_q/\lambda_s)$,
related to conserved quantum numbers:
baryon number and  strangeness, respectively. We also obtain $\lambda_{I3}$
which   expresses 
the asymmetry in the 3-rd component of the isospin. Especially for low 
energy reactions, where the particle yield is relatively low, this parameter differs 
significantly from unity. We have become aware by checking the work of other
groups pursuing statistical hadronization of QGP and fits to hadron yields
that the net charge per net baryon ratio (0.39 for 
heavy nuclei) is not maintained in this work.

%%%%%%%%%%%%%%%%%%%%%%%%%%%%%%%%%%%%%%%%%%%%%%%%%%%%%%%%%%%%%%%%%%%
\section{Hadronization condition}

\begin{figure}[tb] 
\centerline{\psfig{width=9.cm,angle=0,figure= 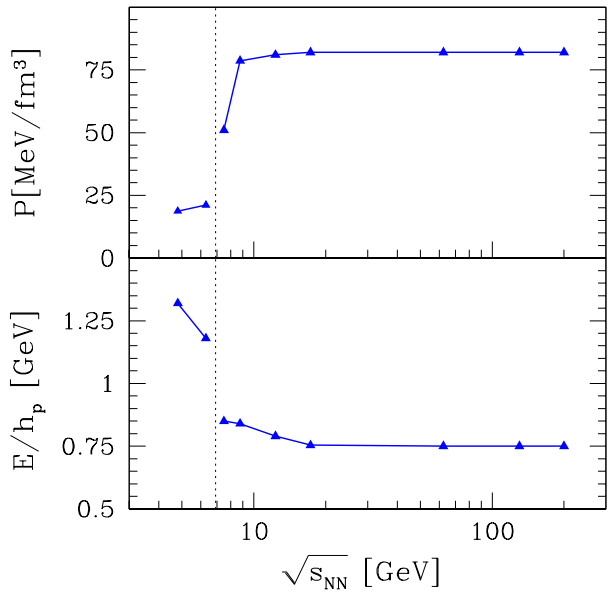}} 
\caption{\label{PEh} Pressure and energy per baryon as function of 
reaction energy. $P$ is fixed for RHIC to the value at SPS to 
help find a reliable reproducible fit.
}
\end{figure}

The SHARE  program provides, beyond statistical parameters, also an opportunity 
to evaluate the physical properties   of the  bulk matter at hadronization.
These  show a change, 
from a low density and low pressure system at low $\sqrt{s}$ (AGS, lowest SPS 20 $A$ GeV data) 
to a highly compressed phase just above this in energy.
In figure~\ref{PEh}, we show, in the upper frame, the pressure $P$ we obtain 
for the different fits. We find~\cite{Letessier:2005qe,Rafelski:2009gu}, 
in the study of the high energy SPS data,  that   hadronization is 
characterized  by a remarkably constant 
value of $P\simeq 82\pm2$ MeV/fm$^3$. This result arises 
in the SPS energy domain (which is involving the total particle
yields) without a   further constrain. For RHIC, the central rapidity results
lack a fixed baryon number and, as shown in previous section, the errors 
are way too large yielding  fit confidence which is too high. 
We thus decided,    at  RHIC, to introduce as 
a additional `measurement'  the value $P\simeq 82\pm2$ MeV/fm$^3$. 

%%%%%%%%%%%%%%%%%%%%%%%%%%%%%%%%%%%%%%%%%%%%%%%%%%%%%%%%%%%%%%%%%%%%%%%Fig 2
%%%%%%%%%%%%%%%%%%%%%%%%%%%%%%%%%%%%%%%%%%%%%%%%%%%%%%%%%%%%%%%%%%%%%%%%%

%%%%%%%%%%%%%%%%%%%%%%%%%%%%%%%%%%%%FIG 3
\begin{figure}[tb] 
\centerline{%\psfig{width=7cm,figure=PLSQRTSPHYSPROPOLD.ps}\hspace*{-.50cm}
\psfig{width=8cm,figure=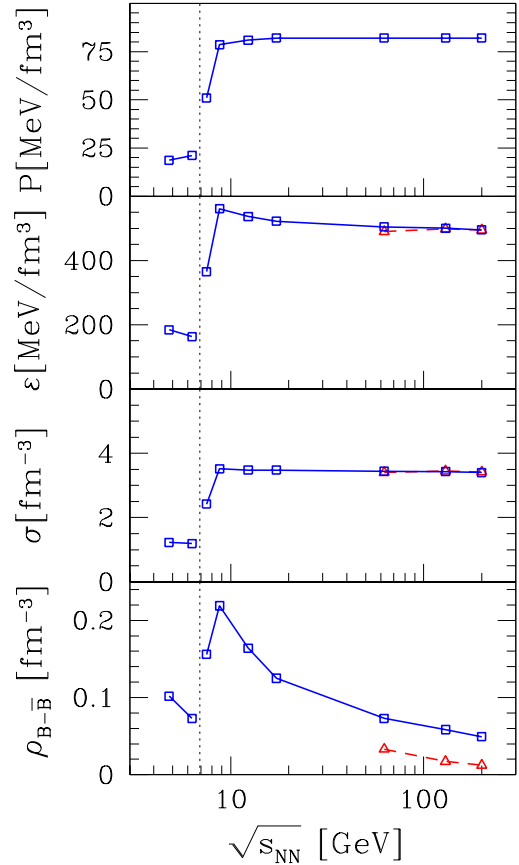}}\vskip -0.3cm
\caption{\label{PHYSPRO}
Physical bulk properties at hadronization beginning with 
highest energy AGS  up to RHIC 200 GeV  results. 
%%%%In red   $dN/dy$   obtained requiring 
Squares (blue) are total 
yield fits, triangle (red) are central rapidity data fits, all 
lines guide the eye, fit results are the symbols. 
}
\end{figure}
%%%%%%%%%%%%%%%%%%%%%%%%%%%%%%%%%%%%%%%%%%%%%%%

In figure \ref{PEh}, we also show in the lower frame the fireball energy per 
primary hadron, $E/h_p$. This value is also remarkably constant for 
top SPS and all RHIC reaction energies. The variation we see is, in 
part, explained by baryon density variation, and at low energy by  
different properties of the hadronizing system also seen in many other
observables. Note that if $\gamma_q^{\rm HG}=1$
is forced, the value $E/h_p\simeq 0.75 $ rises to 1 GeV~\cite{Cleymans:2005xv}. 
Considering the result achieved at SPS we introduce at RHIC as an additional
constraint $E/h_p=0.75\pm10\%$ as is also shown in table \ref{Rhic}. With the 
two constraints, $P$ and $E/h_p$
we find very good data fits with the outcome confirming that  
the hadronization pressure offers a good characterization of QGP breakup. 

Hadron particle pressure emerges, in our study, as a  common physical property 
which defines when and how QGP breaks up into hadrons. Why this is so is 
explained remembering that,  at    $P=1$ atm, water boils in New York and in Beijing at $100^o$\,C.
The color non-conductivity of the true vacuum acts like a `pot cover' keeping quarks 
together, the cover recedes when the pressure is high, QGP expands. After QGP breaks, 
the residual quark pressure turns into hadron pressure. In this picture 
the quark particle pressure
has just the magnitude required to balance the vacuum pressure. Thus, the critical 
pressure of hadronization must be   the vacuum pressure confining color.

The pressure $P$ is compared to several other physical 
bulk properties of QGP in figure~\ref{PHYSPRO}. 
At low energy considering  for example, entropy $\sigma$, 
we note that the AGS and lowest SPS results agree and produce a low 
value, suggesting 
a source which is half as dense compared to other results. This would be 
just what one expects if QGP is not formed at 
low reaction energies, or/and when there has
been a considerable re-equilibration of hadronization products. Consideration 
of other, more penetrating observables, such as strangeness per entropy $s/S$ and 
the continuity of total strangeness production as function of energy support
the second hypothesis, a well equilibrating QGP fireball after hadronization. 

 At  high energy, we see that SPS and RHIC bulk properties  are  consistent. 
Moreover, the  behavior as function of energy of, for example,
the net baryon density (bottom frame) is consistent with the expectation
that it should be decreasing --- since baryon
transparency increasing with energy of reaction 
is an intuitive requirement. Fits which force
$\gamma_i=1$ can fail to produce this natural result. 
 
The values of energy density, 
 $E/V \to 500\,{\rm MeV}\, {\rm fm}^{-3}=(250 \,{\rm MeV})^4$,
and entropy density,  $S/V\to 3.4\, {\rm fm}^{-3}$, obtained 
for the high energy reactions are worth noting. These complement
 $P\to 82\pm2\, {\rm MeV}\, {\rm fm}^{-3}= (158\,{\rm MeV})^4$.

%%%%%%%%%%%%%%%%%%%%%%%%%%%%%%%%%%%%%%%%%%%%%%%%%%%%%%%%%%%%%%%%%%%%%%%%%%%%%%%%%%fig 4
\begin{figure}[htb] 
\centerline{\psfig{width=8.5cm,figure=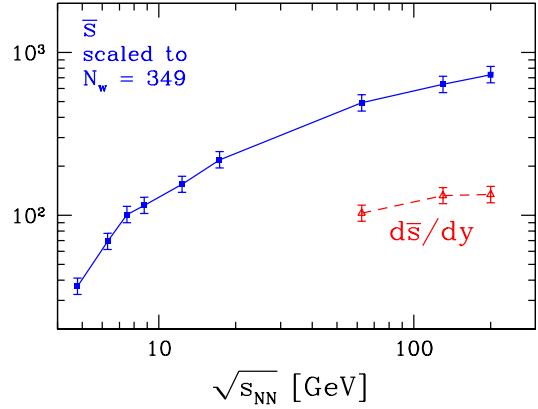}}\vskip -0.7cm
\caption{\label{strange}
Total pair-strangeness yield, triangle (red) for central rapidity at RHIC. }
\end{figure}
%%%%%%%%%%%%%%%%%%%%%%%%%%%%%%%%%%%%%%%%%%%%%%%%%%%%%%%%%%%%%%%%%%%%%%%%%%%%%%%%%%%%%

\section{Particle yield predictions}
In figure~\ref{strange}, we show growth of strangeness pair yield  with energy, 
squares (blue) for the  total hadron most central reaction trigger, 
and RHIC central rapidity most central trigger as triangles (red).
In table \ref{outputRHIC}, we show the particle yields we obtain,
comparison wit table \ref{Rhic} shows that our fit is very 
successful. 
%, including the 
%resonance yields at time of hadronization, which is not necessarily what 
%will be observed, see Ref.~\cite{Kuznetsova:2008hb}.

%%%%%%%%%%%%%%%%%%%%%%%%%%%%%%%%%%%%%%%%%%%%%%%%%%%%%%%%%%%%%%%%%%table 2
\begin{table}[tbh]
\caption{
\label{outputRHIC}
Output hadron multiplicity data for the RHIC energy range. 
See text for the meaning of predictions of $N_{4\pi}$ yields at 
62.4 and 130 GeV and of  $dN/dy$ at 62.4 GeV.
The input statistical parameters are seen in table \ref{Rhic}.
 $b=B-\overline{B}\equiv N_W$ for $4\pi$ results and 
$b=d(B-\overline{B})/dN$ for results at central rapidity.
  Additional significant digits are presented
for purposes of tests and verification. All yields are without the 
weak decay contributions.
}\vspace*{0.2cm}\small   
\begin{center}
\begin{tabular}{|c | c c c |}
\hline
$\sqrt{s_{\rm NN}}$  [GeV]  & 62.4 & 130 & 200  \\
$E_{\rm eq}$[GeV]      &2075 & 9008  &  21321 \\
$\Delta y$  &$\pm 4.2$  & $\pm 4.93$ & $\pm 5.36$ \\
\hline
  & \multicolumn{3}{|c|}{ $dN/dy|_{y=0}$\ 5\% }\\
\hline
$b $                        &35.90 &18.50 &15.1 \\
$\pi^+$                     &232.5 &245.9 &240.7 \\
$\pi^-$                     &235.9 &247.4 &241.9 \\
${\rm K}^+$                 &41.13 &51.9  &50.8 \\
${\rm K}^-$                 &35.14 &48.1  &47.6 \\
${\rm K}_{\rm S}$           &36.99 &48.2  &47.5 \\
$\phi $                     &4.65  &7.54  &7.45 \\
$p $                        &20.86 &15.65 &14.99 \\
$\bar p. $                  &8.43  &10.11 &10.27 \\
$\Lambda$                   &11.4  &11.6  &11.2 \\
$\overline\Lambda$          &5.52  &8.18  &8.26 \\
$\Xi^-$                     &1.71  &2.34  &2.27 \\
$\overline\Xi^+$            &0.99  &1.80  &1.80 \\
$\Omega$                    &0.25  &0.46  &0.45 \\
$\overline\Omega$           &0.17  &0.39  &0.39 \\
%$\Omega+\overline\Omega$   &0.37  &0.69  &0.88 \\
${\rm K}^0(892)  $          &10.7  &12.5  &12.2 \\
\hline
%$\Delta^{0} $               &4.10  &2.98  &2.86 \\
%$\Delta^{++} $              &4.03  &2.97  &2.84 \\
%$\Lambda(1520)$             &0.76  &0.74  &0.72 \\
%$\Sigma^+(1385)$            &1.42  &1.42  & 1.37\\
%$\Xi0(1530) $               &0.58  &0.78  & 0.75\\
%$\eta $                     &26.2  &33.1  & 32.6\\
%$\eta' $                    &2.09  &2.80  & 2.77\\
%$\rho^0 $                   &19.5  &19.2  & 19.0\\
%$\omega(782) $              &17.7  &17.1  & 16.9\\
%$f_0(980)$                  &1.54  &1.47  & 1.45\\
%\hline
%$(s-\bar s)/(s+\bar s)$     &  0   &   0   &  0    \\
%\hline
\end{tabular}\vspace{0.1cm}
\end{center}
 \end{table}
%%%%%%%%%%%%%%%%%%%%%%%%%%%%%%%%%%%%%%%%%%%%%%%%%%%%%%%%%%%%%%%%%%%%%%%%%%%%%%%%%%%%%

\section{Comments and Conclusions} 

We believe that the assumption of $\gamma_q=1$, we often see in literature 
in the context of the analysis of hadron particle yield data,
tests the hypothesis that QGP was {\em not}, or only extremely {\em briefly}
present  in relativistic 
heavy ion collisions. The tacit assumption made is that instead of a QGP, a long-lasting 
 cascade of  hadronic reactions allows the HG particles to chemically equilibrate.
However, if this were to be true, an even stronger evidence for HG dominance of HI reactions would 
be that the value $\gamma_q\to 1$ emerges in the analysis, rather than being assumed.   
However, we find   allowing $\gamma_i\ne 1,\ i=q,\,s$ that, nearly always, the chemical non-equilibrium
prevails, with a much higher confidence level. Furthermore, this additional 
freedom produces QGP like properties of the bulk from which particles emerge.

To summarize, we differ from other groups in the following aspects in our data analysis:
\begin{enumerate}
\item We use  $\gamma_q\ne 1$, and thus allow the ratio of baryon to meson yields 
to be fixed independently of the hadronization temperature;  
\item We enforce, in the fit, 
the conserved ratio of charge to baryon number $Q/b=0.395\pm 0.01$, and are able to fit
the associated $\lambda_{I3}$ fugacity;
\item We do not enforce exact strangeness conservation, 
$\langle s \rangle -  \langle\bar  s \rangle=0$, but instead,  we allow 
$\delta s=(s-\bar s)/(s+\bar s)=0\pm0.05$ to behave like a measurement, the reasoning is as follows:\\
\indent {\bf  a)} Summing all measured and unmeasured hadrons  in strangeness
`conservation' condition, 
$\Delta s=\sum_i h_s^i-\sum_jh_{\bar s}^j\to 0$,  combines  independent  measurement
errors and thus, even if the experimental data had 
all strangeness carrying hadrons, there would
be a residual statistical  error present in  $\delta s$;\\
\indent  {\bf  b)} Some strangeness could escape detection 
in unknown `particles', for example being bound in 
(nearly) $uds$-quark-symmetric semi-stable strangelett (a small drop of quark matter), this
leads to  $\delta s<0$ --- which is what we find as a preferred result  in low energy fits;\\ 
\indent  {\bf  c)} The experiments did  measure  many, but not all relevant
 particles carrying strangeness,   e.g., $\Sigma^\pm$ has not been measured, this yield 
is uncertain and thus $\delta s= 0$ cannot be ever imposed on experimental grounds alone;\\
\indent  {\bf  d)} Introducing an error in $\delta s$, we perform a 
test of the hypothesis that weak decays  remains weak in QGP phase, and 
thus, the net strangeness remains conserved --- another way to understand this is 
to note that  we cannot confirm that weak decays in QGP
remain weak to better than the progressing 
error of individual contributing measurements.
\item In the study of RHIC data, we consider a mix of  STAR and PHENIX particle yield 
results, following the principle that the PHENIX integrated single particle spectra are 
more reliable than those of STAR, and STAR, in turn, given its high acceptance  is more reliable 
in evaluation of yields of hadrons observed in their two or more particle decay channels.
\end{enumerate}
The reader should note that we have, in our fits compared to many other efforts, three more parameters: 
$\gamma_q$ as explicitly stated above, $\lambda_s$ since we do not fix but fit strangeness
conservation and $\lambda_{I3}$ since, as the only group, we enforce a fit to $Q/b$. Thus, we 
have, in general, one less degrees of freedom (three more parameters and two more `data' points,
$\delta s=0\pm0.05$ and $Q/b=0.395\pm 0.01$). 
However, the real issue is that all told we have a 7-dimensional space of parameters, 
$T,\ V, \lambda_q,\ \lambda_s,\ \gamma_q,\ \gamma_s$ and $ \lambda_{I3}$, which 
contains many false minima, and the art of finding the domain of the best fit minimum
is not easily acquired, and cannot be dispensed with the comment `fit is unstable'.
Naturally, it takes much more effort  to find a true minimum in 
a 7-d parameter world, compared to a 2-d parameter world which corresponds to the simplest
and least physical ``equilibrium'' model with $T,\mu_{\rm B}$ as parameters, i.e.,
 setting also $\gamma_s=1$, fixing $\mu_{\rm S}$ by strangeness conservation 
 and using only particle ratios in the fit.  Whoever practices this today has learned
nothing from the work of past 20 years.

The most intriguing result of this analysis is the smoothness, and even near 
constancy, of physical properties of the fireball at chemical freeze-out condition
seen for the top three  SPS energies 40, 80 and 158  $A$GeV, which result agrees 
well with RHIC fits where we impose the pressure $P$ and hadronization particle energy $E/h_p$. 
Of particular physical interest is  the value of hadronization pressure,  $P\simeq 82$ MeV/fm$^3$,
obtained at SPS, and found in this work to be consistent with RHIC data
---  one may imagine that in a phase transformation from quarks to hadrons, the pressure of quarks is 
transferred into the pressure of color-neutral  hadrons, which can escape from the deconfined 
fireball. Since the flow pressure of quarks transfers smoothly into that of hadrons, we conclude that 
the thermal pressure of produced hadrons,  $P\simeq 82$ MeV/fm$^3$, provides a first estimate of 
the pressure of the vacuum which keeps color charged  
quarks inside the fireball up to the point of sudden fireball break-up. 

In summary, we presented a high confidence fit of high centrality data form AGS, SPS and RHIC,
and have found common ground of results at SPS and RHIC regarding the bulk properties 
of hadronizing matter. This suggests that a deconfined phase is with great probability already formed
at or near 30 $A$ GeV.

\ack
Supported  by a grant from the U.S. Department of
Energy,  DE-FG02-04ER41318\,. Laboratoire de Physique Th\'eorique 
et Hautes Energies, LPTHE, at  University Paris 6 and 7 is supported 
by CNRS as Unit\'e Mixte de Recherche, UMR7589.

\end{document}